\documentclass[preprint,12pt]{elsarticle}
\usepackage{amsmath,amssymb,graphicx}
\usepackage{graphicx}
\usepackage{bm}
\usepackage{float}
\usepackage{mathptmx}
\usepackage{epstopdf}
\usepackage{booktabs}
\usepackage{multirow}
\usepackage{hhline}
\usepackage{verbatim}
\usepackage{subfigure}
\usepackage{soul}
\usepackage{blindtext}
\begin{document}


\begin{frontmatter}

\title{Cu-substituted Fe$_2$P: An emerging candidates for magnetic RAM application}
\author{Enamullah} 
\author{S. Bhat} 
\author{Seung-Cheol Lee \corref{cor1}}
\ead{seungcheol.lee@ikst.res.in}
\author{S. Bhattacharjee \corref{cor2}}
\ead{satadeep.bhattacharjee@ikst.res.in}
\cortext[cor1]{Corresponding author}
\cortext[cor2]{Corresponding author}
\address{Indo-Korea Science and Technology Center (IKST), Bangalore 560065, India}

\date{\today}
             
\begin{abstract}
We propose that Cu-substituted Fe$_2$P, (Fe$_{1-x}$Cu$_x$)$_2$P ($x\sim 0.16$), to be an outstanding contender for the STT-MRAM application. Using first principles based calculations in the framework of density functional theory and through Monte Carlo simulations, we demonstrate that this material can be used as ferromagnetic electrode in the magnetic tunnel junction (MTJ) of STT-MRAM due to its moderate perpendicular magnetic anisotropy (PMA), large tunnel magneto-resistance (TMR), good thermal stability and high ferromagnetic transition temperature. We point out that the simplicity in the synthesis, huge abundance, and  non-toxicity make this material a very good candidate to replace the current MTJ materials for STT-MRAM such as FePt,~FeCo or FeCoB.

\end{abstract}

\end{frontmatter}

\section{Introduction}
In the current years, heaps of considerations are given to Spin-transfer-torque magnetic random access memory (STT-MRAM) due to its high density, low power consumption, non-volatility and considered to be one of the promising candidate for next-generation universal memory \cite{Akerman}. The basic unit of a STT-MRAM consists of a magnetic tunnel junction (MTJ) with perpendicular magnetic anisotropy (PMA)~\cite{MRAM1,MRAM2}. The MTJ is constructed with a thin non-magnetic insulating layer sandwiched between two ferromagentic layers. One of the layers has a fixed magnetization while the other layer's magnetization can be rotated ~\cite{rotation1,rotation2}. The bits are recorded in terms of the magneto-resistance of the MTJ. The MTJ has a low resistance if both the ferromagnetic layers have same polarity and usually represents a logical "0" state while the MTJ has a high resistance if the ferromagnetic layers have opposite polarity and is represented by a logical "1" state~\cite{logical1,logical2}. Typically MgO-based MTJ films with ferromagnetic L1$_0$ ordered electrodes such as  FeCo,FeCoB or FePt are used due to their strong interfacial PMA, large tunnel magneto-resistance and thermal stability\cite{material1,material2}.
\par
In the present study, we propose another contender for the ferromagnetic electrode, which can be utilized as a part of MTJ of the STT-MRAM. We show here that (Fe$_{1-x}$Cu$_x$)$_2$P can be used as alternative to FeCo or FePt. This material fulfils  all the essential  requirements to act as component in MTJ. 

Stoichiometric Fe$_2$P is a ferromagnet with Curie temperature (\textit{T$_C$}) of 216 K {\cite{fruchart1969crystallographic}}, exhibiting large Magnetocrystalline anisotropy energy MAE of about 500 $\mu$eV/f.u. {\cite{beckman1991compounds}}. Due to its low \textit{T$_C$}, Fe$_2$P is impractical for room temperature applications. However, earlier studies have shown great enhancement in \textit{T$_C$}, by alloying Fe$_2$P with Ni, Co, Si and B {\cite{fruchart1969crystallographic,beckman1991compounds,chandra1980mossbauer,catalano1973magnetic,jernberg1984mossbauer}}, making them useful for room temperature applications. Interesting magnetic behaviour have been reported for Mn, Cr, Co and Ni substitutions of Fe$_2$P using Mossbauer spectroscopy techniques {\cite{fruchart1969crystallographic,hokabe1974magnetocrystalline,zach2004magneto,kumar2008nature,fujii1978magnetic,jain2007neutron}}. Very small substitutions of Mn (Fe$_{1-x}$Mn$_x$)$_2$P (for $x <$ 0.015) induce metamagnetism and for $x >$ 0.03 it is antiferromagnetic. Similar results are reported for Cr substitution  {\cite{fruchart1969crystallographic,fujii1978magnetic}}. Contrarily, for Co substitution \textit{T$_C$} increases with $x$ to a maximum of 480 K at $x$ = 0.3 {\cite{kumar2008nature}}. Similar behaviour is observed for (Fe$_{1-x}$Ni$_x$)$_2$P compounds with \textit{T$_C$} reaching to maximum (342 K) at about $x$ = 0.1  {\cite{fujii1978magnetic,dolia1993magnetic}}. Also, presence of very small amount of Cu impurities are found to greatly enhance the \textit{T$_C$} {\cite{ericsson1980mossbauer}}.

All the above mentioned studies were intended to make the metal substituted Fe$_2$P to be useful for permanent magnet applications. However no efforts have been made so far to in a direction where non-stoichiometric Fe$_2$P may find a "softer" magnetic applications~\cite{Satadeep,Satadeep1,ref1,ref2} such as switching component in magnetic devices. In the present investigation, we rationalize the attainability of such application.

We explored systematically the the effect of Co, Ni, Cu and Zn substitution on the various magnetic properties of Fe$_2$P, particularly the magnetic moments, Magnetocrystalline anisotropy energy, the ferromagnetic transition temperature. Such screening shows Cu substituted Fe$_2$P is the most suitable candidate for the MRAM application which we have further verified through the calculation of TMR. We discuss our findings following a brief outline of the computational methods.

\section{Computational methods}

We performed first-principles calculations using the projector-augmented wave (PAW) method \cite{blochl1994projector} in the framework of DFT as implemented in VASP code {\cite{kresse1993ab,kresse1996efficient}}. The exchange-correlation energy of electrons is treated within a generalized gradient approximated functional (GGA) of the Perdew−Burke−Ernzerhof (PBE) {\cite{perdew1996generalized}} parameterized form. Interactions between ionic cores and valence electrons are represented using PAW pseudo-potentials, where 4$s$, 3$d$ electrons for transition metals (Fe, Co, Ni, Cu and Zn) and 3$s$, 3$p$ electrons for P are treated as valence. Plane-wave basis set with kinetic energy cutoff of 500 eV and an energy convergence criteria of $10^{-6}$ eV are used. Uniform mesh of 9x9x15 k-points used for Brillouin zone sampling of the unit cell, provided sufficient accuracy. MAE is calculated using force theorem which treats the change in the band energy as a result of the variation of the angle of magnetization axis with respect to the easy axis {\cite{jansen1999magnetic}}.

The Heisenberg exchange coupling constants~$(J_{ij})$ are calculated using spin-polarized relativistic (SPR) Korringa-Kohn-Rostoker (KKR) Green's function method, as implemented within SPRKKR package{\cite{sprkkr}}. Using Heisenberg model, exchange Hamiltonian is given by,

\[
\hat{H} = - \sum_{i \ne j} \hbox{} J_{ij} \hat{e}_{i} \hbox{} \hat{e}_{j}, 
\]

where, $\hat{e}_{i}$ and $\hat{e}_{j}$ are the unit vectors along the direction of local magnetic moment on atomic site $i$ and $j$ respectively. The exchange parameters can be obtained from the energy difference between two different magnetic configurations using the formulation of Liechtenstein et al.\cite{Liech}. An angular momentum cutoff of $l_{max}$=3 and 30 complex energy points was used for the expansion of Green's function. The energy convergence criteria of 10$^{-5}$ is used for self-consistence cycles.  Equilibrium lattice parameters obtained from the {\it{ab}}-{\it{initio}} simulation are used to calculate exchange interaction parameters.  

Once the exchange coupling constants are calculated from SPRKKR package, one can estimate the Curie temperature either using simplified mean filed approximation or via more sophisticated Monte Carlo simulation as implemented in VAMPIRE software package{\cite{vampire1,vampire2}}.
We have used VAMPIRE code to obtain the magnetization versus temperature data and hence the estimation of Curie temperature. The Curie temperature $(T_{C})$ can be obtained using the following expression,
\[
T_{C}  = \frac{\epsilon z J_{ij}}{3 k_{B} }, 
\] 
where, $J_{ij}$ is the pair wise exchange energy either between same or different species of atoms within nearest neighbor approximation, $z$ is the coordination number (number of nearest neighbors), k$_{B}$ is the Boltzmann constant and $\epsilon$ represents a correction factor to account for spin wave fluctuations in different crystal lattices.
\section{Results and Discussions}
Initial structure of Fe$_{2}$P is taken from the experimental data {\cite{carlsson1973determination}} and is optimized by full relaxation of the unit cell and atomic positions. Fe$_2$P crystallizes in hexagonal C22 structure with space group $P\bar{6}2m$ (\#189). The unit cell is composed of three formula units with three Fe atoms (say Fe$_{\text{I}}$) occupy 3$f$, other three Fe atoms (say Fe$_{\text{II}}$) occupy 3$g$, two P atoms occupy 2$c$ (P$_\text{I}$) and one P atom occupies 1$b$ Wyckoff sites (P$_{\text{II}}$). Fe$_\text{I}$ atom is surrounded by four P atoms, whereas Fe$_{\text{II}}$ atom is surrounded by five P atoms, and so referred as tetrahedral and pyramidal sites respectively. The structure can be expressed as Fe3 triangles in the $ab$ plane, with P occupying the alternate layers (see Fig. \ref{fig1}). Computed lattice constants ($a$ = 5.81 and $c$ = 3.43 $\AA$) agree well with the reported experimental values ($a$ = 5.87, and $c$ = 3.46 $\AA$) {\cite{carlsson1973determination}.

An early X-ray diffraction experiment on (Fe$_{1-x}$M$_x$)$_2$P alloys (where M is the transition metal) by Fruchart {\it et al.} {\cite{fruchart1969crystallographic}  revealed Fe$_2$P-type hexagonal structure for $x$ $\leq$ 0.2. Therefore, we took the optimized Fe$_2$P structure and one out of six Fe atoms, either from Fe$_\text{I}$ site or Fe$_{\text{II}}$ site is substituted by transition metal resulting in the formula (Fe$_{1-x}$M$_x$)$_2$P, with $x$ = 0.16 and M = Co, Ni, Cu and Zn.

\subsection{Site preference}
To examine the site preference of the solute M atoms in Fe$_{2-x}$M$_{x}$P, we compared the total energies for both cases, i.e. with M atom at Fe$_\text{I}$ and Fe$_\text{II}$ sites, and the energy differences $\Delta E^{\text {Fe}_{\text{I}}-\text {Fe}_{\text{II}}}$ is listed in Table \ref{table1}. For (Fe$_{1-x}$Co$_x$)$_2$P, total energy is lower by 67 meV/f.u. when Co occupies Fe$_{\text{I}}$ site. This is in agreement with the earlier M$\ddot{o}$ssbauer studies {\cite{fruchart1969crystallographic,kumar2001magnetic}} which disclosed the preferential filling of tetrahedral site. Same trend is obtained for Ni substitution where the energy difference is 21 meV/f.u., which is in accordance with the experimental finding, where Ni atoms occupy Fe$_{\text{I}}$ site preferentially, for $x$ in the range 0 $\leq$ $x$ $\leq$ 0.3, but Fe$_{\text{II}}$ site for $x > 0.7$ {\cite{fruchart1969crystallographic,zach2004magneto,maeda1973mossbauer}}. In case of (Fe$_{1-x}$Cu$_x$)$_2$P and (Fe$_{1-x}$Zn$_x$)$_2$P, our calculations predict that Cu and Zn substitutes Fe preferentially at pyramidal site, for which there are no previous data available for comparison.

\subsection{Magnetic moment}
Table \ref{table2} presents the total and local magnetic moments calculated for 3$f$ and 3$g$ sites for Fe$_2$P and (Fe$_{1-x}$M$_x$)$_2$P alloys. Calculated total magnetic moment (3.01 $\mu_{\text{B}}$/f.u.) for Fe$_2$P agrees well with the experimental value (3.27 $\mu_{\text{B}}$) {\cite{scheerlinck1978neutron}}. The local magnetic moments computed for 3$f$ and 3$g$ sites (0.83 and 2.23 $\mu_{\text{B}}$) are also in accordance with earlier reports (0.96 and 2.31 $\mu_{\text{B}}$/f.u.) {\cite{scheerlinck1978neutron}}. For Co substitution total magnetic moment decreases to 2.71 $\mu_{\text{B}}$/f.u., which can be compared with the experimental value of 2.47 $\mu_{\text{B}}$/f.u. reported for (Fe$_{0.70}$Co$_{0.30}$)$_2$P measured at 12 K {\cite{jain2007neutron}}. For Ni case, computed value is 2.43 $\mu_{\text{B}}$/f.u., which can be compared to experimental value of 2.14 $\mu_{\text{B}}$/f.u. measured at 4 K for (Fe$_{0.75}$Ni$_{0.25}$)$_2$P {\cite{zach2004magneto}}. The agreement between our calculation and experimental values are acceptable as magnetic moments are found to decrease monotonically with increase in $x$, for Co and Ni substitutions {\cite{zach2004magneto,kumar2008nature,fujii1978magnetic}. Calculated $\mu_{\text{total}}$ for Cu and Zn substitution is 2.07 and 1.75 $\mu_{\text{B}}$/f.u., respectively, where Cu has negligible magnetic moment and Zn has negative magnetic moment.

Fig. \ref{fig2} illustrates the variation of MAE and total magnetic moment for stable low energy structures of (Fe$_{1-x}$M$_x$)$_2$P alloys. It is evident form the figure that the total moment decreases linearly as we move from Fe to Zn. Also, MAE decreases for all the transition metal substituted Fe$_{2}$P cases with respect to the pristine alloy. Further, for all the studied cases, decrease in $\mu_{\text{total}}$ is more pronounced when M occupies pyramidal site. That is, substituting at Fe$_{\text{I}}$ site produces little change in magnetic moment, whereas magnetic moment changes significantly for Fe$_{\text{II}}$ substitution.

\subsection{Magnetic anisotropy}
Table \ref{table3} shows the MAE estimated for Fe$_2$P and (Fe$_{1-x}$M$_x$)$_2$P alloys. For Fe$_2$P, computed MAE is 496 $\mu$eV/f.u., which is very close to experimentally determined value of 500 $\mu$eV/f.u. measured at low temperature {\cite{beckman1991compounds}}. Further, our calculation reproduces the observed $c$-axis as the magnetization easy axis. From Table \ref{table3} it is evident that for all M substitutions, the calculated MAE is lower than that of Fe$_2$P. Kumar {\it et al.} {\cite{kumar2008nature}} reported the decrease in MAE for Co substitution up to 10$\%$ in hexagonal phase and then progressive increase with increasing Co in the orthorhombic phase. For Ni substitution, Fujii {\it et al.} {\cite{fujii1978magnetic}} through their experiments revealed a monotonic decrease in MAE with increase in $x$ and dropping to zero at $x$ = 0.3. Thus, our results are in accordance with these experimental findings. For Cu and Zn substitutions there are no previous experimental reports available and our calculations predict their MAE to be 266 and 134 $\mu$eV/f.u., respectively.

Similar to Fe$_2$P, lower energy structures of Co, Cu and Zn substituted Fe$_2$P also have [001] as magnetization easy axis. Since these alloys have \textit{T$_C$} above room temperature and magnetic easy axis oriented along the [001] direction, they can be used in perpendicular magnetic recording applications. For Ni substitution it is reversed, i.e., the magnetic easy axis is along [100] direction. Further, it shows lowest MAE (97 $\mu$eV/f.u.) among all the alloys under consideration (see Fig. \ref{fig2}). Note that the decrease in MAE is substantial ($\sim80\%$) for small substitution of Ni, while the decrease in magnetic moment is marginal ($\sim20\%$). As mentioned earlier, \textit{T$_C$} is maximum for (Fe$_{1-x}$Ni$_x$)$_2$P alloys at $x\sim$0.1, and thus it can be useful for high temperature applications, where high magnetization and low MAE are the desired requirements.
\section{Cu-substituted Fe$_{2}$P as magnetic memory material}
From the above, it is quite evident that the Cu-substituted Fe$_{2}$P satisfies all the necessary requirements. In the following we further investigate its competence as for the same from the calculation of T$_C$ and TMR calculations.
\subsection{Exchange interaction constants and Curie temperature}
Here, we demonstrate the results of interatomic exchange constants for pristine Fe$_{2}$P and Cu-substituted Fe$_{2}$P obtained using SPR-KKR package and the corresponding Curie temperature as presented in Table {\ref{table4}. In order to calculate the Curie temperature accurately beyond the simplified mean field approximation, Monte Carlo simulation is used via VAMPIRE software package\cite{vampire1,vampire2}. For the simulation, we construct a 12$\times$12$\times$7 super cell with periodic boundary condition. To calculate thermal equilibrium magnetization at each temperature, we use 5000 equilibrium time steps. In order to get fast relaxation to thermal equilibrium, the Hinzke-Nowak
combinational algorithm is used in Monte Carlo simulations\cite{Hinzke}. The simulated temperature dependent normalized magnetization plot for Fe$_{2}$P and Cu-substituted Fe$_{2}$P  data are fitted using the Curie-Bloch equation within classical limit\cite{Curie-Bloch},

\[
m(T) = \left( 1- \frac{T}{T_{C}}  \right)^{\beta},
\]
 where, m(T) is normalized magnetization as a function of temperature (T), T$_{C}$ is the Curie temperature and $\beta$  is the critical exponent. In case of Fe$_{2}$P, $\beta$ is 0.501 whereas, for Cu-substituted Fe$_{2}$P it is 0.497. The fitted values of T$_{C}$ for both Fe$_{2}$P and Cu-substituted Fe$_{2}$P are shown in Table{\ref{table3}}. The simulated T$_{C}$ for Fe$_{2}$P is 230K which is in good agreement with the experimental value of 217K, where as in case of Cu-substituted, T$_{C}$ reaches to 792K. We have also calculated the T$_{C}$ with the lower concentration of Cu (8.11\%) and got the value 502K.\\
 \indent To further understand the origin of large T$_{C}$ in the Cu-substituted Fe$_2$P, we show in the Fig.\ref{exchange}, the calculated exchange constants as a function distance. In the topmost panels (a) and (b) we show the calculated exchange constants for pure  Fe$_2$P while in the bottom panels (c), (d) and (e), the exchange constants for the Cu-substituted Fe$_2$P are shown. It can be seen that for the pure case, the strongest interaction is  Fe$_{II}$-Fe$_{II}$ type and is about 14.44 meV. While n the case Cu-substituted Fe$_{2}$P, there are three irons, among which the most prominent interaction is among the F$_{III}$ sites (21.6 meV). It is to be noted that in both the cases, intra-sublattice interaction plays dominant role.

\subsection{Tunnel Magnetoresistance of (Fe$_{1-x}$Cu$_x$)$_2$P (001)/MgO (001)/(Fe$_{1-x}$Cu$_x$)$_2$P(001) trilayer structure}

The magnetic tunnel junction(MTJ) composed of Cu-substituted Fe$_{2}$P having different number of insulating layers of MgO is shown in Fig.\ref{fig-MTJ}. In the figure, different thickness of MgO(001) layers sandwhiches between the four layers of Cu-substituted Fe$_{2}$P(001). Top two and bottom two layers of MTJ have been treated as bulk and remaining layers are fully relaxed using VASP. After getting the spin polarization of both the ferromagnetic slab, the tunnel magnetoresistance(TMR) ratio has been calculated using the following formula,

	\[
TMR = \frac{2P_{L_{1}}P_{L_{2}}}{1-P_{L_{1}}P_{L_{2}}} \times 100\%
\] 

where, $P_{L_{1}}(P_{L_{2}})$ is depicts spin polarization value of ferromagnetic layer, $L_{1}(L_{2})$. The variation of TMR with different number of insulating layer has been shown in Table\ref{table4}. The maximum TMR achieved for 8.11\% of Cu-substituted Fe$_{2}$P having two layers of MgO (94$\%$) which is considerably large and makes the material suitable for STT-MRAM applications.

\section{Summary}
In summary, we have studied the magnetic properties of Fe$_2$P and $M$-substituted Fe$_2$P ($M$ is the transition metal) using first-principles calculations based on DFT. Our calculations show that Co, Ni substitutes Fe preferentially at tetrahedral site, whereas Cu and Zn substitutes at pyramidal site. For all studied alloys, total magnetic moment is less than that of Fe$_2$P and decreases linearly as we move from Co to Zn substitution. Further, decrease in magnetic moment is more pronounced for substitution at pyramidal site. Computed MAE for Fe$_2$P is 496 $\mu$eV/f.u., with magnetic easy axis oriented along the [001] direction, which is in excellent agreement with the experimental results. The MAE of Fe$_2$P is found to decrease for all substitutions. Co, Cu and Zn substituted Fe$_2$P retains [001] as magnetization easy axis, thus can be used in perpendicular magnetic recording applications. Interestingly, Ni substituted Fe$_2$P has [100] as easy axis with lowest MAE of 97 $\mu$eV/f.u. For Cu
-substituted Fe$_2$P, we show that due to its moderate perpendicular magnetic anisotropy, large TMR, it has significant potential to be used as component for a magnetic RAM.
\section*{Acknowledgement}
We acknowledge support from the Convergence Agenda Program (CAP) of the Korea Research Council of Fundamental Science and Technology (KRCF) and Global Knowledge Platform (GKP) program of the Ministry of Science, ICT and Future Planning. Enamullah, as a research associate acknowledges IKST for providing financial support.

\vskip 0.5cm
{\bf References:}
\bibliography{STT}

\clearpage
\newpage
\begin{table}[tb]
	\caption{Calculated total energy difference $\Delta E^{\text {Fe}_{\text{I}}-\text {Fe}_{\text{II}}}$ along with preferred site for M occupation}
\vskip 0.5cm
	\begin{tabular}{lll}
	& {$\Delta E^{\text {Fe}_{\text{I}}-\text {Fe}_{\text{II}}}$ (meV/f.u.)} & {Preferred site} \\
	\hline
	Co & -67 & Fe$_\text{I}$ \\
	Ni & -21 & Fe$_\text{I}$  \\
	Cu & 117 & Fe$_{\text{II}}$  \\
	Zn & 268 & Fe$_{\text{II}}$
	\label{table1}
\end{tabular}
\end{table}
\newpage
\begin{table}
	\caption{Calculated local and total magnetic moment for Fe$_2$P and (Fe$_{1-x}$M$_x$)$_2$P, x=0.16. $\mu_{\text{Fe}_{\text{I}}}$ and $\mu_{\text{Fe}_{\text{II}}}$ are average local magnetic moment at	Fe$_{\text{I}}$ and Fe$_{\text{II}}$ sites, respectively. $\mu_{\text{M}_{\text{I}}}$ and	$\mu_{\text{M}_{\text{II}}}$ represents local magnetic moment of M atom substituted at Fe$_{\text{I}}$ and Fe$_{\text{II}}$ sites, respectively.}
	\vskip 0.5 cm
	\begin{tabular}{llccccc}
	& Site & {$\mu_{\text{Fe}_{\text{I}}}$ ($\mu_{\text{B}}$)} & {$\mu_{\text{M}_{\text{I}}}$ ($\mu_\text{B}$)} & {$\mu_{\text{Fe}_{\text{II}}}$ ($\mu_{\text{B}}$)} & {$\mu_{\text{M}_{\text{II}}}$ ($\mu_{\text{B}}$)} & $\mu_{\text{total}}$ ($\mu_{\text{B}}/$f.u.) \\
	\hline
	Fe$_2$P             & --               & 0.83 & --    & 2.23 & --    & 3.01 \\
	\multirow{2}{*}{(Fe$_{1-x}$Co$_x$)$_2$P} & Fe$_{\text{I}}$  & 0.75 & 0.37  & 2.13 & --    & 2.71 \\
	                    & Fe$_{\text{II}}$ & 0.88 & --    & 2.30 & 0.87  & 2.66 \\
	\multirow{2}{*}{(Fe$_{1-x}$Ni$_x$)$_2$P} & Fe$_{\text{I}}$  & 0.59 & 0.12  & 2.03 & --    & 2.43 \\
	                    & Fe$_{\text{II}}$ & 0.88 & --    & 2.18 & 0.28  & 2.39 \\
	\multirow{2}{*}{(Fe$_{1-x}$Cu$_x$)$_2$P} & Fe$_{\text{I}}$  & 0.42 & 0.004 & 1.90 & --    & 2.15 \\
	                    & Fe$_{\text{II}}$ & 0.82 & --    & 1.93 & 0.04  & 2.07 \\
	\multirow{2}{*}{(Fe$_{1-x}$Zn$_x$)$_2$P} & Fe$_{\text{I}}$  & 0.27 & -0.04 & 1.75 & --    & 1.88 \\
	                    & Fe$_{\text{II}}$ & 0.77 & --    & 1.53 & -0.12 & 1.75
	\label{table2}
\end{tabular}
\end{table}
\begin{table}[tb]
	\caption{Calculated MAE along with magnetic easy and hard axis for Fe$_2$P and (Fe$_{1-x}$M$_x$)$_2$P with x=0.16}
	\vskip 0.5cm
	\begin{tabular}{llccc}
	      & Site & Easy Axis & Hard Axis & MAE ($\mu$eV/f.u.) \\
	\hline
	Fe$_2$P             & --               & 001 & 100 & 496 \\
	\multirow{2}{*}{(Fe$_{1-x}$Co$_x$)$_2$P} & Fe$_{\text{I}}$  & 001 & 100 & 199 \\
	                    & Fe$_{\text{II}}$ & 001 & 100 & 188 \\
	\multirow{2}{*}{(Fe$_{1-x}$Ni$_x$)$_2$P} & Fe$_{\text{I}}$  & 100 & 010 & 97  \\
	                    & Fe$_{\text{II}}$ & 001 & 100 & 146 \\
	\multirow{2}{*}{(Fe$_{1-x}$Cu$_x$)$_2$P} & Fe$_{\text{I}}$  & 100 & 001 & 266 \\
	                    & Fe$_{\text{II}}$ & 001 & 100 & 182 \\
	\multirow{2}{*}{(Fe$_{1-x}$Zn$_x$)$_2$P} & Fe$_{\text{I}}$  & 100 & 001 & 134 \\
	                    & Fe$_{\text{II}}$ & 001 & 100 & 157 
	\label{table3}
\end{tabular}
\end{table}
\newpage

\begin{table}
\caption{Calculated exchange interaction constants ($J_{ij}$ in meV/link) and Curie temperature ($T_{C}$ in $K$) for Fe$_2$P and Cu-substituted Fe$_{2}$P. $J_{ij}$'s are considered only between Fe atoms. In Fe$_{2}$P, there are two inequivalent sites for Fe-atom, whereas, in Fe$_{2-x}$Cu$_{x}$P ($x$=16.67\%), there are three inequivalent sites. The indices, I,II and III represent FeI, FeII and FeIII atoms respectively.}
\vskip 0.5cm
	\begin{tabular}{llcccccc}
	System & $J_{I-I}$ & $J_{I-II}$ & $J_{I-III}$ & $J_{II-II}$ & $J_{II-III}$ & $J_{III-III}$ & $T_{C}$ \\
	\hline
	Fe$_2$P             & 5.13 & 9.39 & -- & 14.44 & -- & -- & 230\\
	Fe$_{2-x}$Cu$_{x}$P & 9.84 & 1.18 & 8.10 & -0.05 & 3.41 & 21.60 & 792 
	\label{table4}
\end{tabular}
\end{table}
\begin{table}
\caption{TMR for Cu-substituted Fe$_{2}$P MTJ with different number of MgO insulating layers.}
\vskip 0.5cm
	\begin{tabular}{llccc}
	layer & TMR for $x$=8.33$\%$ & TMR for $x$=16.67$\%$  \\
	\hline
	2 layer & 94$\%$ & 58$\%$  \\
	4 layer & 13$\%$ & 74 $\%$ \\
	6 layer & 2$\%$ & 15$\%$ &
	\label{table4}
\end{tabular}
\end{table}

\newpage

\begin{figure}
	\centering	 
	\includegraphics[scale=0.5]{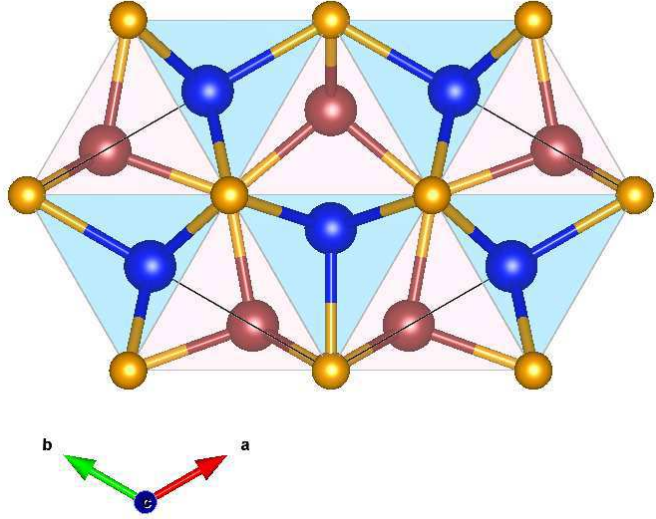}
	\caption{\label{fig1} (color online). Fe$_2$P crystal structure. Fe$_{\text{I}}$ tetrahedral sites (brown), Fe$_{\text{II}}$ pyramidal sites (blue) and P atoms (yellow).}
\end{figure}
\clearpage

\begin{figure}
	\centering	 
	\includegraphics[scale=0.5]{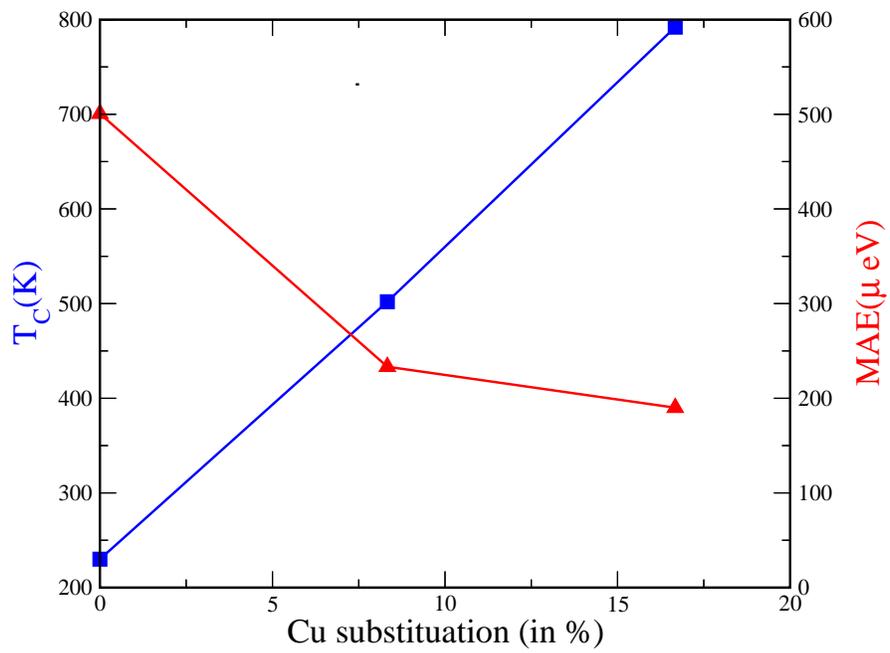}
	\caption{\label{fig_TC_mea}(color online) Variation of Curie temperature (T$_{C}$) and MAE versus Cu concentration $x$. Plot depicts that increasing the doping concentration increases T$_{C}$ and decreases MAE.}
\end{figure}
\clearpage

\begin{figure}
	\centering	 
	\includegraphics[width=10cm,height=10cm]{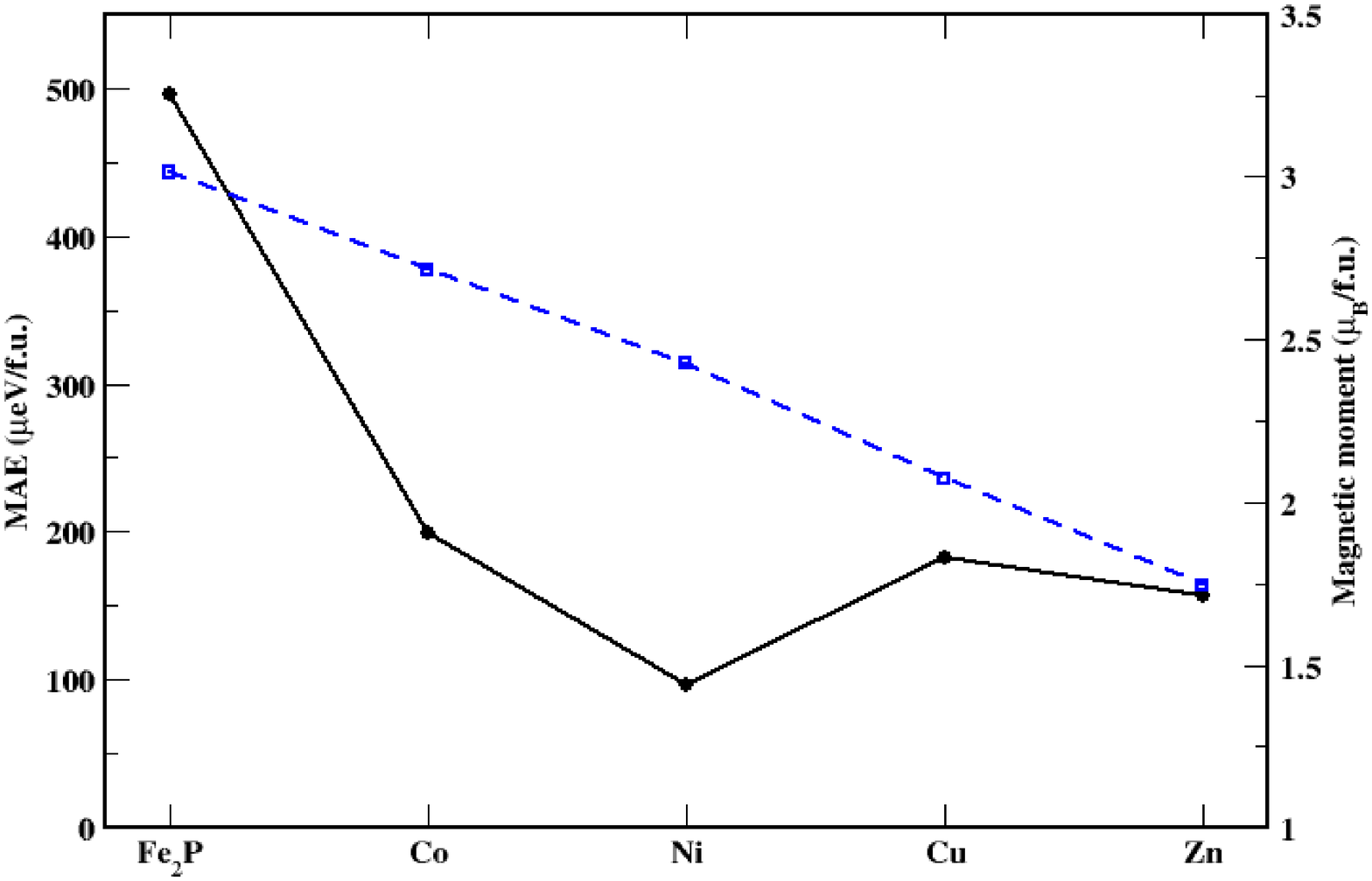}
	\caption{\label{fig2} (color online). Calculated MAE and magnetic moment for pristine and Co, Ni, Cu and Zn substituted Fe$_2$P. Dashed and solid lines represent magnetic moment and MAE respectively.}
\end{figure}
\clearpage
\newpage
\begin{figure}
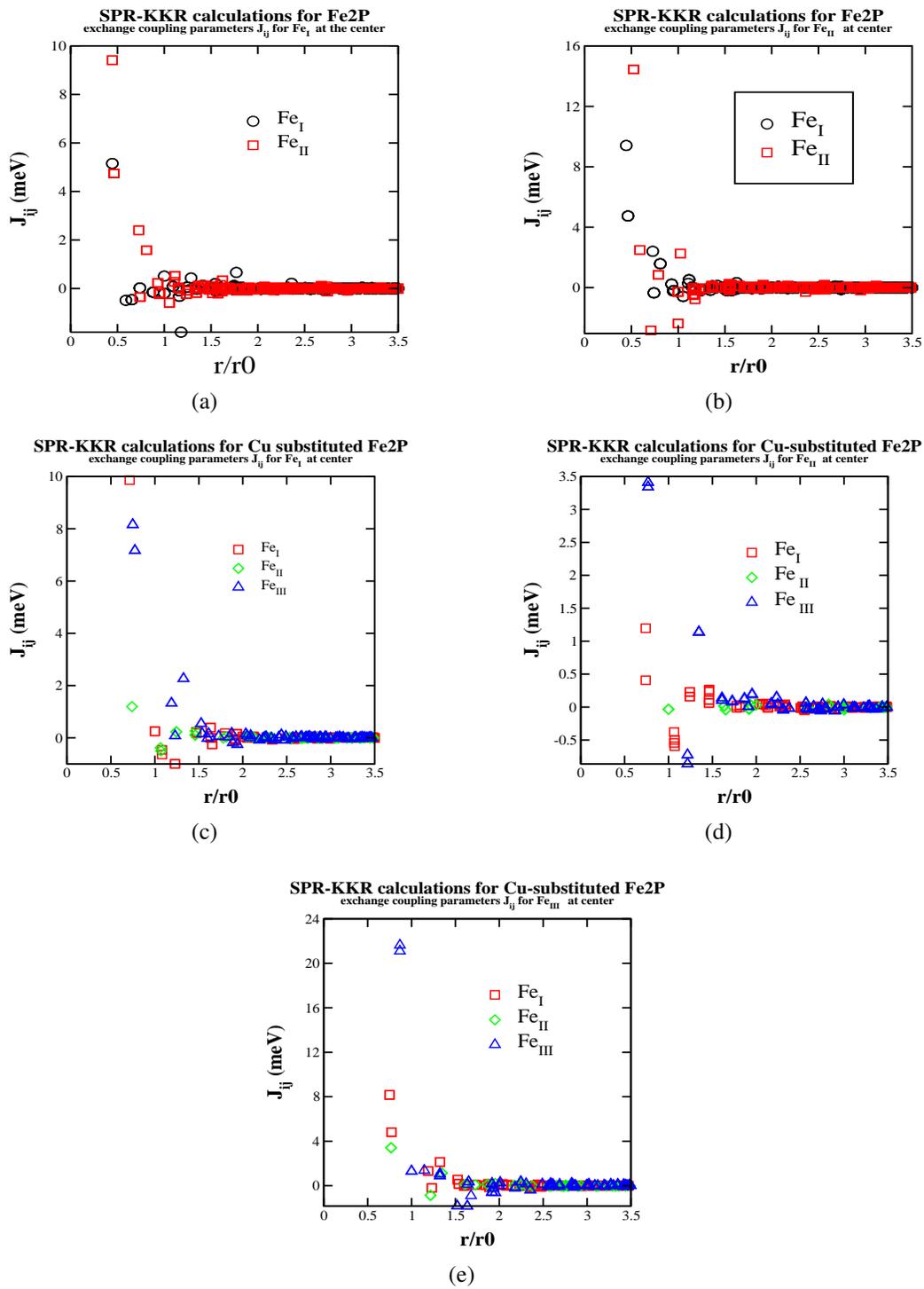

\subfigure[]{\includegraphics[width=60mm,height=55mm]{Fe2P-1.eps}}
\subfigure[]{\includegraphics[width=60mm,height=55mm]{Fe2P-2.eps}}
\subfigure[]{\includegraphics[width=60mm,height=55mm]{S1.eps}}
\hfill \subfigure[]{\includegraphics[width=60mm,height=55mm]{S2.eps}}
\center
\subfigure[]{\includegraphics[width=60mm,height=55mm]{S3.eps}}
\caption{Calculated exchange constants for the pure Fe$_2$P  and Cu-substituted Fe$_2$P }
\label{exchange}
\end{figure}
\clearpage
\begin{figure}
	\centering	 
	\includegraphics[scale=0.5]{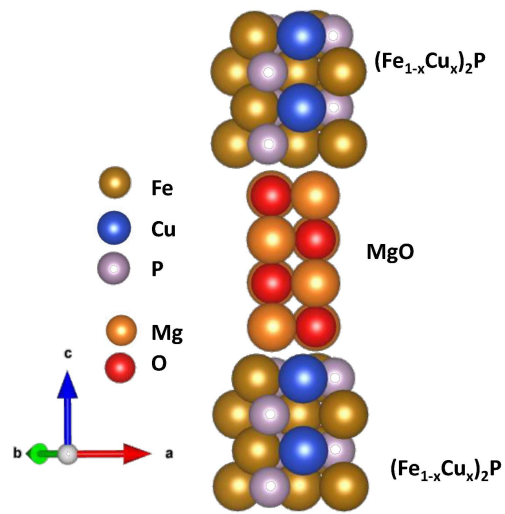}
	\caption{\label{fig-MTJ}(color online) Magnetic Tunnel Junction (MTJ) with Cu-substituted Fe$_2$P and MgO}
\end{figure}
\clearpage

\end{document}